\documentclass{emulateapj}
\usepackage{apjfonts}
\newcommand{\tnu}{\tilde{\nu}}
\newcommand{\tgamma}{\tilde{\gamma}}
\newcommand{\tkappa}{\tilde{\kappa}}
\newcommand{\ttau}{\tilde{\tau}}
\newcommand{\teta}{\tilde{\eta}}
\slugcomment{accepted to ApJ Letters}
\shorttitle{Radio polarization of GRB afterglows}
\shortauthors{Toma et al.}
\begin{document}
\title{
Probing the Efficiency of 
Electron-Proton Coupling in Relativistic Collisionless
Shocks through the Radio Polarimetry of Gamma-Ray Burst Afterglows
}
\author{Kenji Toma\altaffilmark{1}, Kunihito Ioka\altaffilmark{1,2}, 
and Takashi Nakamura\altaffilmark{1}}

\altaffiltext{1}{Department of Physics, Kyoto University,
Kyoto 606-8502, Japan}
\altaffiltext{2}{Theory Division, KEK (High Energy Accelerator
Research Organization) 1-1 Oho, Tsukuba 305-0801, Japan}
\email{toma@tap.scphys.kyoto-u.ac.jp}
\begin{abstract}
The late-time optical/radio afterglows of $\gamma$-ray bursts 
(GRBs) are believed to be synchrotron emission of electrons 
accelerated in relativistic collisionless shocks propagating in 
the ambient medium of the sources.
However, the fraction $f$ of electrons that are coupled to protons
and accelerated remains unclear and a large number of thermal
electrons that are not coupled to protons may be left behind.
If $f<1$, the true explosion energies of GRBs are $f^{-1}$ times
larger than those commonly estimated with $f=1$.
Thus the value of $f$ gives an important constraint on the nature 
of the central engine of GRBs and the physics of collisionless shocks.
Although early-time radio observations can probe the thermal 
electrons, they are difficult at present.
We show that the Faraday rotation effects of the thermal electrons
may suppress the linear polarization of the afterglow at
frequencies higher than the absorption frequency in the late time, 
if the magnetic field is ordered at least in parts, and that
$f$ can be constrained through the observation of the effects.
We find that those effects may be detected with late-time, 
$\geq 1$~day, polarimetry with ALMA for a burst occurring 
within 1~Gpc (i.e., $z \simeq 0.2$), if $f \sim 10^{-1}$.
\end{abstract}


\section{Introduction}
\label{sec:intro}
The afterglows of $\gamma$-ray bursts (GRBs) have been observed
mainly in the optical and radio wavebands in the late time
(i.e., several hours after the burst trigger) since the late 1990's,
and they are widely explained as 
due to synchrotron emission of electrons accelerated in relativistic 
collisionless shocks driven into the ambient medium of the GRB sources.
The synchrotron emission mechanism is supported by the detection of 
linear polarization at the level of $\sim 1-3\%$ in several optical
afterglows 
\citep[for reviews,][]{covino04,lazzati06}.
These understandings are also being confirmed by recent observations 
with \textit{Swift} satellite, although the situation is very 
complicated in the X-ray band and in the early time 
\citep[for reviews,][]{piran04,meszaros06,zhang07}.

In the standard external shock model of GRB afterglows, the late-time
dynamics of the shock is determined by the explosion energy $E$ and the 
ambient medium number density $n$.
We usually treat the fractions of the explosion energy that go into
the magnetic field and electrons and the fraction of electrons that
gain the energy of protons as free parameters
$\epsilon_B, \epsilon_e,$ and $f$, respectively, since
these have not been derived from the basic principles.
We additionally assume that 
all the electrons that gain the proton energy are
accelerated to form a power-law energy distribution, 
$dn/d\gamma_e \propto \gamma_e^{-p}$ 
for $\gamma_e \geq \gamma_m \simeq \epsilon_e(m_p/m_e)\Gamma$,
where $\Gamma$ is the Lorentz factor of the shocked fluid,
and a thermal component with $\gamma_e \simeq \epsilon_e(m_p/m_e)\Gamma$
is not produced.
We commonly constrain the parameters $\{E,~n,~\epsilon_B,~\epsilon_e,~p\}$
by the observations, assuming $f=1$ \citep[e.g.,][]{panaitescu02,yost03}.

This implies that current observations cannot constrain 
the electron-proton coupling parameter $f$.
\citet{eichler05} showed that the observations also allow
the external shock model with the parameters chosen as
$\{E'=E/f,~n'=n/f,~\epsilon_B'=\epsilon_B f,~\epsilon_e'=\epsilon_e f, 
~p'=p\}$ in which the fraction $f$ of total electrons gains the 
proton energy and the fraction $(1-f)$ is thermal electrons with 
$\gamma_e \simeq \Gamma$, as long as $m_e/m_p < f < 1$.
(Hereafter we call the former and latter electron components
``accelerated electrons" and ``thermal electrons", respectively.)
The parameter $f$ gives an important constraint on the true
explosion energies of GRBs.
It also gives a clue to unveil the physics of collisionless shocks.

The thermal electrons provide additional emission and absorption for
the afterglow flux only for $\nu < \tnu_m$,
where $\tnu_m$ is the characteristic synchrotron frequency of the thermal 
electrons.
This results in the sharp decline or sharp rise of flux as $\tnu_m$ 
passes the observed frequency, which can be detected through
early-time radio observations,
$t \lesssim 10^3$~s for $\nu \gtrsim 10^{11}$~Hz \citep{eichler05}.
However, such observations are difficult at present.
Furthermore, it is unclear whether the standard model is applicable for
$t \lesssim 10^3$~s \citep[see e.g.,][]{ioka06,toma06}.

In this paper, we show that the thermal electrons give the Faraday
effects on the afterglow polarization substantially even in the late time 
and that $f$ can be constrained through the observation of the effects.
Those effects may be significant (even for $\nu > \tnu_m$) if the magnetic
field is ordered to some extent \citep{matsu03,sagiv04}.
The frequency below which the Faraday rotation effect is significant 
is expected to be orders of magnitude higher 
than that supposed so far with $f=1$,
because the position angle rotation depends on electron Lorentz factor
as $\Delta\chi \propto (\ln\gamma_e)/\gamma_e^2$.

\section{Transfer of Polarized Radiation}
\label{sec:transfer}
The transfer of polarized radiation through spatially homogeneous
plasma with a weakly anisotropic dielectric tensor may be described
by the transfer equation of Stokes parameters
\citep{sazonov69,jones77,melrose80a,melrose80b},
\begin{equation}
\left(
\begin{array}{cccc}
d/ds + \kappa_I & \kappa_Q & 0 & \kappa_V \\
\kappa_Q & d/ds + \kappa_I & \kappa_V^* & 0 \\
0 & -\kappa_V^* & d/ds + \kappa_I & \kappa_Q^* \\
\kappa_V & 0 & -\kappa_Q^* & d/ds + \kappa_I
\end{array}
\right)
\left( 
\begin{array}{c}
I \\ Q \\ U \\ V 
\end{array}
\right)
=
\left(
\begin{array}{c}
\eta_I \\ \eta_Q \\ 0 \\ \eta_V 
\end{array}
\right),
\label{eq:transfer}
\end{equation}
where 
$s$ is a length parameter along the ray path, and
the right-handed system of coordinates with the wavevector
$\mathbf{k}$ along axis 3 and the magnetic field
$\mathbf{B}$ on plane 2-3 is adopted.
Here $\eta_{I,Q,V}$ are polarization-dependent emissivity, and
$\kappa_{I,Q,V}$ ($\kappa_{Q,V}^*$) are the transfer coefficients 
related to the anti-Hermitian (Hermitian) part of the dielectric tensor, 
describing polarization-dependent absorption
(the polarization of the normal modes of the plasma).
If $|\kappa_V^*| \gg |\kappa_Q^*|$, the normal modes are 
circularly polarized, and the transfer equation (\ref{eq:transfer})
indicates that the conversion of $Q$ and $U$ occurs.
This is well-known Faraday rotation.
If $|\kappa_Q^*| \gg |\kappa_V^*|$, the normal modes are linearly 
polarized and the conversion of $U$ and $V$ occurs.
This is called Faraday conversion.

We define the optical depth $\tau = \kappa_I s$, the rotation depth
$\tau_V = |\kappa_V^*|s$, and the conversion depth $\tau_Q =
|\kappa_Q^*|s$.
The properties of the solution of the transfer equation 
(\ref{eq:transfer}) are as follows.
First, suppose that the absorption effect is not significant, i.e., 
$\tau \ll 1$.
In this case the equation (\ref{eq:transfer}) may be integrated 
easily \citep{melrose80b,jones77}.
For $|\kappa_V^*| \gg |\kappa_Q^*|$, we obtain the linear 
polarization
\begin{equation}
\Pi_L = \frac{\sqrt{Q^2 + U^2}}{I} \simeq 
\frac{\eta_Q}{\eta_I} \left|\frac{\sin(\tau_V/2)}{\tau_V/2}\right|,
\label{eq:faraday}
\end{equation}
and the circular polarization is given by the intrinsic one, 
$\Pi_C = |V|/I = |\eta_V/\eta_I|$.
For $\tau_V \gg 1$, the linear polarization damps.
This results from the fact that the emission from different points 
through the source have its polarization plane rotated at different
angles.
Analogously, for $|\kappa_Q^*| \gg |\kappa_V^*|$ and $\tau_Q \gg 1$,
the damping of $\Pi_C$ occurs while $\Pi_L$ remains intrinsic.

Secondly, in the case in which the absorption effect is significant, 
i.e., $\tau \gg 1$, we may obtain the polarization degrees
approximately by eliminating the differential term from equation 
(\ref{eq:transfer}).
As an example, if the Faraday effects are weaker than the absorption
effect, i.e., $\kappa_I^2 \gg {\kappa_V^*}^2$ and $\kappa_I^2 \gg
{\kappa_Q^*}^2$, 
\begin{equation}
\Pi_L \simeq 
\left|\frac{\eta_Q/\eta_I - \kappa_Q/\kappa_I}
{1-(\eta_Q/\eta_I)(\kappa_Q/\kappa_I)}\right|
\label{eq:absorption}
\end{equation}
is obtained to the leading order.
The circular polarization is similarly given by 
$\simeq |\eta_V/\eta_I - \kappa_V/\kappa_I|$.

In the following sections, we apply this formulation to the 
late-time GRB afterglows.
The anisotropic part of the dielectric tensor is tens of magnitude
smaller than unity for the shocked plasma of a typical GRB afterglow.
We assume that
(1) the pitch-angle distribution of electrons is isotropic for
simplicity;
(2) the shocked plasma is spatially homogeneous;
\footnote{
Electron cooling makes the electron energy distribution inhomogeneous,
but it can be neglected in the late phase of the afterglow 
\citep{sari98}.
}
(3) the shocked plasma consists of a number of random cells within
each of which magnetic field is ordered.
With the third assumption, we obtain the observed linear and circular 
polarizations by $1/\sqrt{N}$ times those for completely
ordered magnetic field, where
$N$ is the number of the random cells in the visible region
\citep{jones77,gruzinov99}.
To reproduce the optical detection at the level of $\sim 1-3\%$
\citep{covino04}, $N$ would be $\sim 10^3$.

\section{Polarization of late-time GRB afterglows}
\label{sec:application}
In this section, we derive the polarization spectrum of
the late-time afterglow, based on the standard external shock model
in which all the electrons are accelerated, i.e., $f=1$
(see \S~1).
The energy distribution of the electrons is assumed to be
$dn/d\gamma_e \propto \gamma_e^{-p}$ for $\gamma_e \geq \gamma_m$.
The transfer coefficients for such electron plasma 
are summarized for frequency region
$\nu > \nu_m$ by \citet{jones77} and for $\nu_B \ll \nu \ll \nu_m$
by \citet{matsu03},
\footnote{
We adopt the value of 
$\kappa_{\alpha}^{* Q} = (\alpha + \frac{3}{2})/2$ 
different from that shown in \citet{jones77}, and the sign of 
$\kappa_V$ should be changed in \citet{matsu03}.
} where $\nu_m$ is the characteristic 
synchrotron frequency corresponding to $\gamma_m$ 
and $\nu_B$ is the nonrelativistic electron Larmor frequency.

The radius of the shock and the Lorentz factor of the shocked fluid 
evolve as $R \simeq (17 Et/4\pi m_p c n)^{1/4}$ and 
$\Gamma \simeq (17 E/ 1024 \pi m_p c^5 n t^3)^{1/8}$, respectively,
where $t$ is the observer time \citep{sari98}.
The comoving width of the shocked plasma shell can be estimated by
$R/4\Gamma$, which we use as the path length of the transfer equation
(\ref{eq:transfer}).
The magnetic field strength, the minimum Lorentz factor and number
density of the accelerated electrons are written by
$B=(32\pi m_p c^2 \epsilon_B n)^{1/2} \Gamma$, 
$\gamma_m = \epsilon_e (m_p/m_e) \Gamma$, and
$n_{\rm{acc}} = 4\Gamma n$, respectively.
Then we obtain
$\nu_B \simeq 4\times10^6 E_{52}^{1/4} n_0^{1/4} \epsilon_{B,-2}^{1/2}
t_d^{-3/4}$~Hz and 
$\nu_m \simeq 6\times10^{12} E_{52}^{1/2} \epsilon_{B,-2}^{1/2}
\epsilon_{e,-1}^2 t_d^{-3/2}$~Hz, respectively.
Here (and hereafter) we have adopted the notation $Q_x=Q/10^x$ in cgs
units and $t_d = t/1$~day.

\begin{figure}
\epsscale{1.1}
\plotone{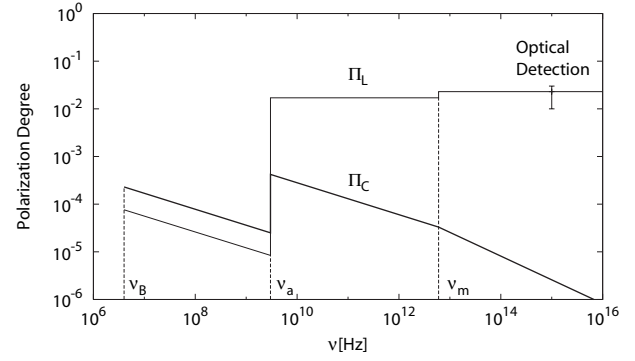}
\caption{
Polarization spectrum of a late-time GRB afterglow, at $t=1$~day,
inferred from the standard external shock model in which all the 
electrons are accelerated, i.e., $f=1$.
The degrees of linear polarization $\Pi_L$ (thin solid line) and
circular polarization $\Pi_C$ (thick solid line) are shown.
The degrees are calculated as $1/\sqrt{N} \sim 10^{-1.5}$ times 
those for completely ordered magnetic field, 
i.e., they are calibrated by detected optical linear 
polarization \citep{covino04}.
Typical values of parameters are used: 
$E=10^{52}$~ergs, $n=1~{\rm cm}^{-3}$, $\epsilon_B = 10^{-2}$,
$\epsilon_e = 10^{-1}$, and $p=2.2$. 
}
\label{fig1}
\end{figure}

Figure~\ref{fig1} illustrates the polarization spectrum of the late-time
GRB afterglow.
The frequencies at which $\tau, \tau_V,$ and $\tau_Q$ equal unity are 
given by
$\nu_a \simeq 3\times10^{9} E_{52}^{1/5} n_0^{3/5} \epsilon_{B,-2}^{1/5}
\epsilon_{e,-1}^{-1}$~Hz,
$\nu_V \simeq 10^9 E_{52}^{3/16} n_0^{9/16} \epsilon_{B,-2}^{1/4}
\epsilon_{e,-1}^{-1} t_d^{-1/16}$~Hz,
and
$\nu_Q \simeq 10^9 E_{52}^{1/5} n_0^{3/5} \epsilon_{B,-2}^{1/5}
\epsilon_{e,-1}^{-1}$~Hz,
where $p=2.2$ has been used as a fiducial value.
Since $\nu_a > \nu_Q \simeq \nu_V$, so that no plasma effects are
significant in the optically thin regime $\nu > \nu_a$ and the intrinsic 
degree of polarization is obtained,
$\Pi_L = \eta_Q / \eta_I = 0.5$ and 
$\Pi_C = |\eta_V / \eta_I| 
\simeq \gamma_m^{-1} (\nu/\nu_m)^{-1/3}$ for $\nu \ll \nu_m$.
For $\nu > \nu_m$, $\Pi_L = (p+1)/(p+\frac{7}{3}) \simeq 0.7$ and
$\Pi_C \simeq \gamma_m^{-1} (\nu/\nu_m)^{-1/2}$.
In the optically thick regime $\nu < \nu_a$, $\tau^2 \gg \tau_V^2$ and
$\tau^2 \gg \tau_Q^2$ are satisfied, the linear polarization is given by
equation (\ref{eq:absorption}).
Because $\eta_Q/\eta_I = \kappa_Q/\kappa_I = 0.5$ for $\nu \ll \nu_m$,
the intrinsic linear polarization vanishes and $\Pi_L$ is only produced
by the conversion of the circular polarization.
The transfer equation (\ref{eq:transfer}) indicates that  
$\Pi_L \approx (\kappa_Q^*/\kappa_I)(\eta_V/\eta_I - \kappa_V/\kappa_I) 
\approx 2\times10^{-2} |\eta_V/\eta_I|$ and
$\Pi_C \approx \eta_V/\eta_I - \kappa_V/\kappa_I 
\approx 6\times10^{-2} |\eta_V/\eta_I|$.
All the characteristic frequencies $\nu_a, \nu_Q$, and $\nu_V$ are
weakly dependent of time, so that the polarization spectrum does not
evolve significantly.
The suppression of $\Pi_C$ due to absorption effect has not been pointed
out in the context of GRB afterglows, since \citet{matsu03} erroneously used 
$\kappa_V$ of opposite sign and \citet{sagiv04} neglected $\kappa_V$.
\citet{sagiv04} discussed similar propagation effects in early-time
afterglows, but their discussion should be restricted to a frequency
region $\nu \gg \nu_B (\simeq 10^{10}-10^{11}~\textrm{Hz})$.
For $\nu \gg \nu_B (> \nu_a)$, $\Pi_L$ and $\Pi_C$ they derived for the 
forward shock emissions are consistent with our results.

\section{A signature of the thermal electrons}
\label{sec:signature}
Here we derive the polarization spectrum 
according to the standard external shock model in which the thermal
electrons are left behind, i.e., $f<1$ 
(see \S~1), and
show that the linear polarization may be suppressed even at frequencies
higher than the absorption frequency $\nu_a$.
The electron energy distribution is assumed to consist of the 
accelerated electrons which are considered 
in \S~3
and the thermal electrons with the Lorentz factor $\tgamma_m = \Gamma$
and the number density $n_{\rm{th}} = [(1-f)/f]n_{\rm{acc}}$,
where $m_e/m_p < f < 1$
\citep[see Fig.~1 of][]{eichler05}.
Then all the quantities in this model can be written by using 
the parameters $\{E,~n,~\epsilon_B,~\epsilon_e,~p\}$ as measured 
assuming $f=1$, while the real values of the parameters are given by
$\{E'=E/f,~n'=n/f,~\epsilon_B'=\epsilon_B f,~\epsilon_e'=\epsilon_e f, 
~p'=p\}$.
The characteristic synchrotron frequency of the thermal electrons
is estimated by
$\tnu_m \simeq 2\times10^8 E_{52}^{1/2} \epsilon_{B,-2}^{1/2}
t_d^{-3/2}$~Hz.
We approximate the transfer coefficients for the thermal electrons
as those for the monoenergetic distribution of electrons
\citep{sazonov69,melrose80a,melrose80b}.
Thus we consider the electron energy distribution
\begin{equation}
\frac{dn}{d\gamma_e} = n_{\rm{th}} \delta(\gamma_e - \tgamma_m)
+ K \gamma_e^{-p} H(\gamma_e - \gamma_m),
\end{equation}
where $K = (p-1)n_{\rm{acc}}\gamma_m^{p-1}$ and $H(x)$ is the 
Heaviside step function.
(Hereafter we describe the quantities related to the thermal electrons
as $\tilde{Q}$.)
For $\nu \gg \tnu_m$, $\teta_{I,Q,V}$ and $\tkappa_{I,Q,V}$ damp 
exponentially.
The remaining coefficients for the thermal electrons are different 
from those for the power-law distribution only by numerical factors.
Here we show the expressions of Faraday coefficients, 
\begin{equation}
\tkappa_V^* \simeq \frac{1}{\pi} \frac{e^2}{m_e c} n_{\rm{th}}
  (2\pi \nu_B \cos\theta) \tgamma_m^{-2} (\ln\tgamma_m) \nu^{-2},
\label{eq:kappav} 
\end{equation}
\begin{equation}
\tkappa_Q^* \simeq 
\left\{
\begin{array}{ll}
  \frac{2^{1/3} \pi^{1/3}}{3^{11/6} \Gamma_{\rm{E}}(\frac{1}{3})} 
  \frac{e^2}{m_e c}
  n_{\rm{th}} (2\pi \nu_B \sin\theta)^{2/3} \tgamma_m^{-5/3} \nu^{-5/3}
  & \textrm{for}~~ \nu \ll \tnu_m, \\
  -\frac{1}{2\pi^2} \frac{e^2}{m_e c} n_{\rm{th}} (2\pi\nu_B \sin\theta)^2
  \tgamma_m \nu^{-3} 
  & \textrm{for}~~ \nu \gg \tnu_m,
\end{array} \right.
\end{equation}
where $\theta$ is the angle between $\mathbf{k}$ and 
$\mathbf{B}$ and $\Gamma_{\rm{E}}(x)$ is the Euler gamma function.
The coefficients for the electron energy distribution consisting of
the thermal plus accelerated ones are given by the linear combination
of the two contributions.

\begin{figure}
\epsscale{1.1}
\plotone{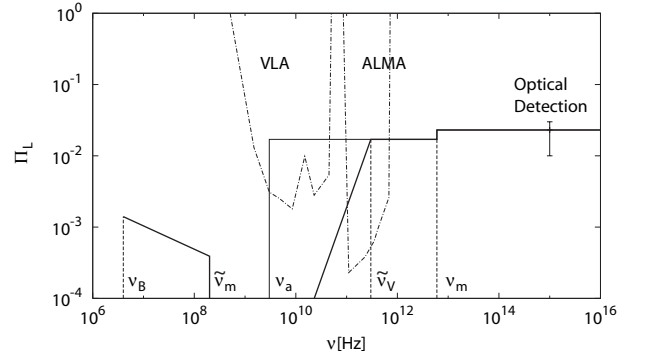}
\caption{
Linear polarization spectra of a late-time GRB afterglow, at
$t=1$~day, inferred from the standard external shock models in
which all the electrons are accelerated (thin solid line;
the same of $\Pi_L$ in Figure~\ref{fig1}) and in which
the thermal electrons are left behind with 
$(1-f)/f = 10$ (thick solid line).
The parameters $E,~n,~\epsilon_B,~\epsilon_e,~p,$
and $N$ are the same as Figure~\ref{fig1}.
The dot-dashed lines describe the sensitivities of VLA and ALMA
for a burst with $D = 1$~Gpc ($z \simeq 0.2$) and an integration 
time of 1 hour.
At frequencies lower than $\tnu_V$, $\Pi_L$
is suppressed by the Faraday rotation effect of the thermal 
electrons, which can be detected with ALMA.
}
\label{fig2}
\end{figure}

In Figure~\ref{fig2}, we show
the linear polarization spectrum of the late-time GRB afterglow for 
the $f<1$ model, compared with that for 
the $f=1$ model obtained 
in \S~3.
The sensitivities of ALMA and VLA for 1 hour integration time
are also shown.
They are derived by $\Pi_L \geq F_{s,\nu}/F_{\nu}$, where 
$F_{s,\nu}$ is the sensitivity limit for
continuum radiation and $F_{\nu}$ is the flux of the afterglow.
The flux is estimated by $F_{\nu} = F_m(\nu/\nu_m)^{1/3}$ for 
$\nu_a<\nu<\nu_m$ and $F_{\nu} = F_m(\nu_a/\nu_m)^{1/3}(\nu/\nu_a)^2$ for
$\tnu_m<\nu<\nu_a$, where $F_m \simeq 10^2 D_{27.5}^{-2} E_{52} n_0^{1/2}
\epsilon_{B,-2}^{1/2}$~mJy and $D$ is the luminosity distance
\citep{sari98}.

For $\nu \gg \tnu_m$, the absorption effect of the thermal electrons
is absent, and thus the absorption frequency 
is the same as the $f=1$ case.
Since $\ttau_V/\tau_V \simeq [(1-f)/f](\tgamma_m/\gamma_m)^{-2}
(\ln\tgamma_m/\ln\gamma_m) \gg 1$
(see equation (\ref{eq:kappav}))
and similarly $\ttau_Q/\tau_Q \gg 1$, the Faraday effects are 
dominated by those of the thermal electrons.
The ratio $\ttau_V/\ttau_Q \simeq \tgamma_m^{-1} (\ln\tgamma_m)
(\nu/\tnu_m)$ is $\gg 1$ for small $\tgamma_m$, i.e., at the late phase of
the afterglow, so that the normal modes of this plasma is 
circularly polarized and the Faraday rotation effect is significant.
The frequencies at which $\tau_V$ and $\tau_Q$ equal unity
are given by
\begin{equation}
\tnu_V \simeq 3\times10^{11} \left[\frac{(1-f)/f}{10}\right]^{1/2} 
E_{52}^{3/16} n_0^{9/16} \epsilon_{B,-2}^{1/4} t_d^{-1/16}
~\textrm{Hz}, 
\label{eq:rotation}
\end{equation}
and
$\tnu_Q \simeq 4\times10^{10} [(1-f)/10f]^{1/3} 
E_{52}^{1/3} n_0^{1/3} \epsilon_{B,-2}^{1/3} t_d^{-2/3}$~Hz,
respectively.
For $\nu > \tnu_V$, all the depths are smaller than unity, so that
the intrinsic polarization is obtained.
In the regime $\nu_a < \nu < \tnu_V$, 
$\ttau_V \gg \ttau_Q \gg 1 \gg \tau$ is satisfied, so that $\Pi_L$
is given by equation (\ref{eq:faraday}).
It damps at low frequencies as $\propto \nu^2$ and oscillates with 
the period $|\Delta\nu/\nu| \sim 10^{-1} \nu_{11}^2$.
In the optically thick regime $\tnu_m \ll \nu < \nu_a$,
the transfer equation (\ref{eq:transfer}) for $\ttau_V \gg \ttau_Q
\gg \tau \gg 1$ indicates that 
$\Pi_L \approx (\ttau_Q/\ttau_V)(\eta_V/\eta_I - \kappa_V/\kappa_I)$
and
$\Pi_C \approx \eta_V/\eta_I - \kappa_V/\kappa_I$ \citep{jones77},
and thus $\Pi_L$ does not exceed $\Pi_C$.
For $\nu \ll \tnu_m$, both the absorption and the Faraday effects are
dominated by the thermal electrons, and $\ttau^2 \gg \ttau_V^2$
and $\ttau^2 \gg \ttau_Q^2$ are satisfied.
Then the polarization spectrum is similar to that for $\nu<\nu_a$
in the $f=1$ model discussed 
in \S~3.
It is important to note that both $\Pi_L$ and $\Pi_C$ are $<10^{-2}$
for $\nu < \nu_a$, and they are far from detectable because the flux
is suppressed in this regime 
(especially for $\nu < \tnu_m$, 
the additional absorption by the thermal electrons exists).

The existence of the thermal electrons is characterized by the 
suppression of the linear polarization at $\nu_a < \nu < \tnu_V$.
Necessary conditions for this suppression
are $\tnu_V \gg \nu_a$ and $\tnu_V \gg \tnu_Q$.
The former condition reduces to 
$(1-f)/f \gg 10^{-3} E_{52}^{1/40} n_0^{3/40}
\epsilon_{B,-2}^{-1/10} \epsilon_{e,-1}^{-2} t_d^{1/8}$.
Interestingly, the effect can be seen even for as small number of 
thermal electrons as $(1-f)/f \sim 10^{-1}$. 
The latter condition implies that
$t \gg 3\times10^3 [(1-f)/f]_1^{-8/29}
E_{52}^{7/29} n_0^{-11/29} \epsilon_{B,-2}^{4/29}$~s,
which shows that \textit{the late-time afterglow is suitable to search for 
the existence of the thermal electrons through the observation of 
linear polarization.}

If $\nu_a$ is determined by the observation of a bright burst and
the linear polarization is not detected at $\nu \gtrsim \nu_a$ with VLA
and detected at $\nu \gg \nu_a$ with ALMA, it becomes clear that a number 
of the thermal electrons exist and
the magnetic field is ordered on large scales.
If we determine $\tnu_V$,
the electron-proton coupling parameter $f$
can be constrained by equation (\ref{eq:rotation}).

\section{Discussion}
\label{sec:discussion}
We have studied a signature of the thermal electrons only in the patchy
coherent magnetic field model, while there are some other viable models
for magnetic field configuration.
In the model of random field with very short coherence length 
\citep[e.g.,][]{sari99,ghisellini99}, the coefficient $\tkappa_V^*$
averaged over the field configuration vanishes, so that the Faraday
depolarization of $\Pi_L$ does not occur \citep{matsu03}.
In the model of a combination of random field $B_{\rm rnd}$ and 
large-scale ordered field $B_{\rm ord}$ \citep{granot03},
the depolarization by $B_{\rm ord}$ can occur similarly as discussed
in \S~4.
In this model $\Pi_L \simeq 0.7 B_{\rm{ord}}^{1.6}/
\langle B_{\rm{rnd}}^{1.6} \rangle$ for $\nu > \nu_m$
and $p=2.2$, so that $B_{\rm{ord}}^{1.6}/
\langle B_{\rm{rnd}}^{1.6} \rangle \sim 10^{-1.5}$
to reproduce the optical detection.
Interestingly for $\tnu_V < \nu \ll \nu_m$, 
$\Pi_L = 0.5 B_{\rm{ord}}^{2/3}/ \langle B_{\rm{rnd}}^{2/3}
\rangle \sim 0.1$.
If such a high $\Pi_L$ at $\tnu_V < \nu \ll \nu_m$ is detected,
it will be an evidence for the presence of $B_{\rm ord}$.

Only upper limits have been obtained so far for the radio polarization 
from GRB afterglows \citep{granot05}.
From bright GRB 030329, $\Pi_L \simeq 2\%$ is measured in the optical band
\citep{greiner03},
whereas $3 \sigma$ limits $< 1 \%$ are derived at 8.4~GHz.
Such a low degree at radio may be attributed to the source being optically 
thick, since $\nu_a$ is estimated as $\simeq 19$~GHz \citep{taylor05}.

If a large number of thermal electrons are left behind, i.e., $f<1$,
the afterglow energy of GRBs should be $E'=Ef^{-1}$, where $E$ is the 
afterglow energy estimated by $f=1$ model and typically inferred to be 
$\sim 10^{51.5}$~ergs with jet collimation correction 
\citep{panaitescu02,yost03}.
The association of GRBs with supernovae suggests that
$f > 10^{-1.5}$ is a conservative lower limit.
The energy of prompt $\gamma$-ray emission is typically similar to $E$,
and some of the models of early-time afterglows imply that
the efficiency of the $\gamma$-ray emission is
$\gtrsim 90\%$ \citep[e.g.,][]{ioka06,toma06,granot06,fan06}.
If the external shock model with $f<1$ is applicable to early-time 
afterglows, the $\gamma$-ray efficiency problem would be solved.

\citet{mundell07} have reported a $2 \sigma$ upper limit
$< 8\%$ on the optical polarization in early-time afterglow of 
GRB 060418 ($t\sim 200$~s), and argued that presence of a large-scale
ordered field in the GRB jet is ruled out.
However, the rotation frequency $\nu_V$ for a typical reverse shocked 
ejecta with ordered field is $\sim 10^{15}$~Hz, so that the low level 
of polarization degree would result from the Faraday depolarization 
\citep[see][]{sagiv04}. 

It is suggested that there are electrons well coupled to protons at
some nonrelativistic collisionless shocks, that is, 
$\epsilon_e \approx 0.5$ \citep[e.g.,][]{markevitch06}.
However, the fraction $f$ of total electrons that are coupled to 
protons has not been discussed seriously.
Recently, \citet{spit07} has reported the results of 2-dimensional
particle-in-cell simulations of relativistic collisionless shocks in
electron-proton plasma with realistic value of $m_p/m_e$ that
$\epsilon_e \approx 0.5$ and $f \approx 1$ were realized.
However, it is too early to interpret the results conclusive, since
long-term 3-dimensional simulations with good resolution have not been 
done and it has not been understood whether the heated electrons are
accelerated into the power-law energy spectrum.
The late-time radio polarimetry may be an important test for more 
realistic simulations and theories.

\acknowledgements
We thank N.~Kawanaka and K.~Murase for useful comments.
This work is supported 
by the Grant-in-Aid 18740147 (K.~I.),
19047004 (T.~N.), and 19540283 (T.~N.) from the MEXT of Japan.

\end{document}